# Remark on "Algorithm 916: Computing the Faddeyeva and Voigt functions": Efficiency Improvements and Fortran Translation

MOFREH R. ZAGHLOUL, United Arab Emirates University[1]

This remark describes efficiency improvements to Algorithm 916 [Zaghloul and Ali 2011]. It is shown that the execution time required by the algorithm, when run at its highest accuracy, may be improved by more than a factor of two. A better accuracy vs efficiency trade off scheme is also implemented; this requires the user to supply the number of significant figures desired in the computed values as an extra input argument to the function. Using this trade-off, it is shown that the efficiency of the algorithm may be further improved significantly while maintaining reasonably accurate and safe results that are free of the pitfalls and complete loss of accuracy seen in other competitive techniques.

The current version of the code is provided in *Matlab* and *Scilab* in addition to a *Fortran* translation prepared to meet the needs of real-world problems where very large numbers of function evaluations would require the use of a compiled language. To fulfill this last requirement, a recently proposed reformed version of Humlíček's w4 routine, shown to maintain the claimed accuracy of the algorithm over a wide and fine grid is implemented in the present *Fortran* translation for the case of 4 significant figures. This latter modification assures the reliability of the code to be employed in the solution of practical problems requiring numerous evaluation of the function for applications tolerating low accuracy computations ($<10^{-4}$).

---

Author's addresses: M. Zaghloul, Department of Physics, College of Sciences, United Arab Emirates University, Al-Ain, 15551, UAE.

## 1. INTRODUCTION

Many algorithms have been proposed in the literature to evaluate the Faddeyeva function, $w(z)$ (defined mathematically as the scaled complementary error function for a complex variable),

$$w(z) = e^{-z^2}(1 - \text{erf}(-iz)) = e^{-z^2}\text{erfc}(-iz) \qquad (1)$$

where $i = \sqrt{-1}$, $z = x+iy$ is a complex argument, and erfc($z$) is the complementary error function. The Voigt function, $V(x,y)$, used in spectroscopy and astronomy is the real part of the Faddeyeva function. Algorithm 916 [Zaghloul and Ali 2011] is a recent valuable addition to the existing set of Faddeyeva function evaluation algorithms.

The most important advantage of Algorithm 916 is its high accuracy compared to other competitive algorithms. In particular, Algorithm 916 does not suffer from the catastrophic failure (complete loss of accuracy) near the real axis, which occurs with some methods employing rational approximations. In some practical applications, evaluation of the Faddeyeva function may account for a significant fraction of the execution time; for example, in the line-by-line calculation of microwave and infra-red radiative transfer from spectroscopic data (for more details, see [Letchworth and Benner 2007]). For such applications it is important to have a version of the code that may be compiled, i.e., implemented in *Fortran* rather than *Matlab*.

In Section 2 we provide details of the changes we made in order to improve the efficiency of the function evaluation for high accuracy computations. A more effective accuracy vs efficiency trade-off scheme is explained in Section 3 where we also show that an additional efficiency gain by a factor of three may be obtained while still maintaining safe results of reasonable accuracy. A *Fortran* translation of the code is described in Section 4 and benchmarking results are provided in Section 5.

## 2. HIGH ACCURACY COMPUTATIONS

A set of efficiency improvements has been implemented in the present version of the code, *Faddeyeva_v2.m*. The main changes are summarized as follows:
1. In the original implementation of the function [Zaghloul and Ali 2011] the series summations ($\Sigma_1$, $\Sigma_2$, $\Sigma_3$, $\Sigma_4$, and $\Sigma_5$) are performed using a single computational loop. Numerical experiments were performed to ascertain the minimum number of loop cycles needed for the convergence of all the series to the highest possible accuracy for a wide range of input values. Results using IEEE double precision floating point arithmetic showed that using 13 loop-cycles was always sufficient to reproduce the highest possible accuracy obtained by Algorithm 916. We have, therefore, replaced the convergence check procedures of the original code by a fixed number of 13 loop-cycles which results in a reasonable saving of CPU time.
2. The remaining exponential terms inside the computational loop used for the summation of all series are no longer calculated inside the loop. They are either pre-calculated in an array using the fact that the calculations are now performed for a fixed number of loop cycles or replaced by products using recurrence relations where the intrinsic exponential function is replaced by a multiplication process inside the loop. This also results in a saving of execution time.

3. The array indices for $|x|=0$, and for $|x| \geq \sqrt{-\ln(R_{min})}$, where $R_{min}$ is the smallest positive normalized floating point number, and while $|z|<|z|_{border}$ are identified using the *Matlab find* function. The conditional loop used for the calculation of the sums in the original version of the algorithm is replaced by two independent loops in addition to a few simple array expressions. (The identification of $|z|_{border}$ is explained in Section 3.) Avoiding unnecessary repetitions of some computations and using the highly efficient array computations for others reduces execution time.

4. For $|x|=0$ we use $w(0+iy) = \text{erfcx}(y)$ and for large values of $|z|$ we keep the third convergent of the Laplace continued fractions, which approximates $w(z)$ asymptotically [Faddeeva and Terent'ev 1961, Abramowitz and Stegun 1965, Gautschi 1970],

$$w(z) \cong \frac{i}{\sqrt{\pi}} \frac{1}{z-} \frac{\frac{1}{2}}{z-} \cdot \frac{1}{z-} \ldots \approx \frac{i}{\sqrt{\pi}} \frac{2(z^2-1)}{(2z^3-3z)} \tag{2}$$

The accuracy of this asymptotic expression increases as $|z|$ increases. Hence, we carefully implement this expression with a moving boundary matching the required accuracy as explained below. Using this asymptotic expression for the region $|z|^2 \geq |z|^2_{border} = 3.8 \times 10^4$ reproduces the highest accuracy results from Algorithm 916, in its range of applicability. The maximum relative error for both the real, $V$, and imaginary, $L$, parts ($\delta_V = |(V-V_{ref})|/V_{ref}$ and $\delta_L = |(L-L_{ref})|/L_{ref}$) is less than $10^{-13}$ when using Algorithm 680 [Poppe and Wijers. 1990] as a reference for this region.

5. Some unnecessary calculations are dropped from the present version of the code. For example, the scaled complementary error function for real numbers, erfcx(y), is now calculated only for values of the vector y corresponding to , as it is not required otherwise. This saves both memory and CPU time.

In addition to the above changes, many other less effective changes and coding improvements have been implemented in the present version of the code. For completeness, derivatives of the Faddeyeva function are also calculated and returned.
For the array of 2,840,710 points generated using the grid *y=logspace(-20,4,71)* and *x=linspace(-200,200,40001)* used in [Zaghloul and Ali 2011] and on the same machine, the run time has been reduced from $\approx 4.46$ s for the old version of the code (run for the highest possible accuracy) to $\approx 1.55$ s for the present version of the code when run for the highest accuracy (see Table 1 in Section 3 below). Calculations have been performed using *Matlab 7.9.0.529 (R2009b)*.

## 3. ACCURACY VS EFFICIENCY TRADE-OFF

Despite its superior accuracy, Algorithm 916 is competitive or sometimes faster than other highly accurate techniques in the region of small values of $|z|$, e.g. $|z| \leq 6$. However, for large values of |z|, the performance is slower. It has to be noted that in cases where the function is used as an approximation, accuracy of the order of $10^{-4}$ (the same accuracy as the supplied input data) is sufficient or at least acceptable.

Reducing the number of computational loop-cycles used in the calculation of the series summation, to a number less than 13, marginally improves efficiency at the expense of accuracy. On the other hand, moving the boundary of applicability of the asymptotic approximation to smaller values of |z| in line with the required accuracy further improves the efficiency of computation. It may be helpful to recall here that the maximum absolute relative error of using the approximation in (2) for both of the real and imaginary parts decreases as |z| increases and vice versa. Accordingly, we have two free parameters that affect both the accuracy and the execution time; the number of loop-cycles, *Ncycles*, and the chosen boundary at which to apply (2), *bigz_border*$\equiv |z|^2_{border}$. By fixing *Ncycles* at 13 and changing the value of *bigz_border*, the dependence of the accuracy on the position of the border may be determined as shown by the first two columns in Table 1. Similarly, we may vary the value of *Ncycles* while keeping the value of *bigz_border* fixed at $3.8\times10^4$ to determine the dependence of the accuracy on the number of computational cycles as shown by the first and third columns of Table 1. The run times, in seconds, obtained by using the `best' combination of *bigz_border* (second column) and *Ncycles* (third column) for each of the possible user specified accuracies (first column) is given in the fourth column in Table 1.

**Table 1** Accuracy vs efficiency trade-off showing the average runtime (4$^{th}$ column) corresponding to the desired accuracy (1$^{st}$ column) using the `best' combination of *bigz_border* (2$^{nd}$ column) and *Ncycles* (3$^{rd}$ column), as adopted in the present version of the code. Average run time has been calculated for the 2,840,071 points generated from the grid *y=logspace(-20,4,71)* and *x=linspace(-200,200,40001)* and using *Matlab 7.9.0.529 (R2009b)*.

| Number of significant figures "sdgts" | Ncycles=13 | bigz_border $\|z\|^2_{border}=3.8\times10^4$ | Average run time (s) |
|---|---|---|---|
| | "bigz_border" $\|z\|^2_{border}$ | "Ncycles" | |
| 13 | $3.80\times10^4$ | 13 | 1.550 |
| 12 | $1.75\times10^4$ | 12 | 1.200 |
| 11 | $8.20\times10^3$ | 11 | 0.964 |
| 10 | $3.80\times10^3$ | 11 | 0.870 |
| 9 | $1.75\times10^3$ | 10 | 0.759 |
| 8 | 810 | 10 | 0.704 |
| 7 | 380 | 9 | 0.604 |
| 6 | 180 | 9 | 0.547 |
| 5 | 110 | 8 | 0.491 |
| 4 | 107 | 7 | 0.491 |

## 4. A FORTRAN VERSION

Algorithm 916 was originally implemented as a *Matlab* function. However, some practical problems require a large number of evaluations of the Faddeyeva function ($10^{11}$ or more) [Letchworth and Benner 2007]. Such large scale computations are invariably performed using compiled languages rather than packages like *Matlab* or any of its clones (*Octave* or *Scilab*). Hence, a *Fortran* translation of the current version of the code has been implemented. The translation is presented in the form of a module, *Faddeyeva_v2_mod_rk*, which contains an *elemental* subroutine *Faddeyeva_v2_rk(z, sdgts, w, dVdx, dVdy, Stat)*. The routine can be run using single or double precision arithmetic depending on the choice of the integer parameter *rk* in the subsidiary module *set_rk*. The elemental subroutine, *Faddeyeva_v2_rk*, operates on a single dummy complex variable, *z*, while the desired number of significant figures in the computed results *sdgts* is provided as an additional *intent (in)* argument. In addition to the returned values of the corresponding Faddeyeva function, *w*, partial derivatives of the real part of the Faddeyeva function *dVdx* and *dVdy* are returned as optional outputs. An additional optional argument, *Stat* may be used to test the exit status of the function; 0 signifies a successful calculation, 1 is returned if *sdgts* <4 on entry, 2 is returned if *sdgts* > 13 or 6 for the double or single precision version of the function and, finally, -1 is returned if an overflow occurs during the computation. The routine may be invoked with arrays as actual arguments where it will be applied element-wise, with a conforming array return value.

The recommended range for *sdgts* is between 4 and 13 for double and between 4 and 6 for single precision computations. Use of a value of *sdgts* less than 4 is not recommended for accuracy concerns, particularly regarding the computations of the derivatives, and such values will be replaced by 4. User supplied values of *sdgts* > 13 or 6 (double or single precision respectively) are not recommended for performance reasons and such values will be reset to 13 or 6 respectively.

The subroutine uses Cody's algorithm for calculating the scaled complementary error function, *erfcx*, for real numbers (under the name "*erfc_scaled*") [Cody 1969]. The *Fortran* 2008 standard [ISO/IEC:2010 ITP ] defines a new intrinsic function, *erfc_scaled*, although this is not currently implemented by many *Fortran* compilers. To use this new intrinsic the *USE rk_erfcx_Cody* statement in the *Faddeyeva_v2_mod_rk* module needs to be commented out.

## 5. SPEED BENCHMARKING
### 5.1. Matlab Speed Benchmarking

The present version of the function, *Faddeyeva_v2.m*, allows two input arguments; the array of the complex numbers, *z*, and the desired level of accuracy in the calculated Faddeyeva function expressed in terms of the integer number of significant figures "sdgts". When "sdgts" is not provided, it is directly set to a default value of 4, which may be considered the lowest acceptable accuracy for reliable calculations of the derivatives. An accuracy requirement of 4 significant figures requires 7 computational loop-cycles.

Speed benchmarks are sensitive to the actual environment and conditions and are influenced by numerous factors. In particular, the *tic* and *toc* functions in *Matlab* measure the overall elapsed time. Thus it is necessary that no other applications are running in the background of the computational system that could affect the timing of the *Matlab* programs. Warming up the function to be timed is required in order to eliminate one-time effects related to file systems, memory systems, M-file parsing and optimization, etc., which could contribute to inconsistent timings sometimes seen on the first run. To reduce first-run timing effects on speed benchmarking results we often generate several consecutive times and use the best time obtained.

Recent versions of *Matlab* (e.g. *R2013b)* include a new timing function, *timeit*, which is also available on the *Matlab* File Exchange [Eddins 2010]. The function *timeit* runs the code several times to ensure that any first-run timing effects do not affect the results and returns the *median time*.

We have used the function *timeit* for the *Matlab* performance tests for all versions of Algorithm 916 and the results are summarized in Table 2. However, because of the relatively long run time of the *Scilab* version of the code, the function was only run a few times and the average run time is presented for comparison.

**Table 2** Comparison between average runtime of different *Matlab* and *Scilab* versions of Algorithm 916 (including calculation of the derivatives) for the 2,840,071 points generated using the grid *y=logspace*(-20,4,71) and *x=linspace*(-200,200,40001). Calculations have been performed using *Matlab 7.9.0.529 (R2009b)*, *Scilab 5.4.1*. Reference values are taken from Algorithm 680 [Poppe and Wijers. 1990] except for the region in the vicinity of *x*=6.3 and small values of *y* where values of the function calculated using erfi(*z*) from *Mathematica*$^{TM}$ are used.

| Code version | | Max. of $\delta_V$ & $\delta_L$* | Run time (s) | |
|---|---|---|---|---|
| | | | *Matlab* | *Scilab* |
| Original Version | *Faddeyeva(z, 1.43e-17)* 13 significant digits | ~$10^{-13}$ | 4.46 | 1110.2 |
| | *Faddeyeva(z)* 4 significant digits | ~$10^{-4}$ | 2.70 | 459.6 |
| Modified Current Version | *Faddeyeva_v2(z, 13)* 13 significant digits | ~$10^{-13}$ | 1.550 | 588.3 |
| | *Faddeyeva_v2(z, 4)* 4 significant digits | ~$10^{-4}$ | 0.491 | 33.6 |

* $\delta_V = |(V - V_{ref})|/V_{ref}$ and $\delta_L = |(L - L_{ref})|/L_{ref}$.

The results shown in Table 2 confirm the improvement of the execution time of the new version of the code by a factor close to 3 when run at the highest accuracy. The results also illustrate the effectiveness of the newly implemented trade off scheme where a further reduction of the execution time by a factor close to 3 is obtained for the grid under consideration.

## 5.2. Fortran Speed Benchmarking

It is also useful to compare the execution time of the present *Fortran* version with other algorithms available in the literature. Three algorithms (available to us in *Fortran*) have been used to conduct the comparison; Algorithm 680 [Poppe & Wijers 1990] which is known for its claimed high accuracy $\sim 10^{-13}$, Humlíček's algorithm [Humlíček 1982] which is known for its remarkable efficiency and claimed low accuracy $<10^{-4}$ and Weideman's algorithm [Weideman 1994] with a claimed accuracy better than $10^{-6}$ when using $N=16$ terms and accuracy of 12 significant digits when using $N=32$ terms.

Four different cases have been worked out to perform such a speed comparison. For all worked cases, a grid of 2,840,071 input values for $z$ is used with 71 values for $y$ and 40001 values for $x$. Table 3 summarizes the details of these four run cases while the results of the comparison are given in Table 4. A comparison between the present *Fortran* version when run using single precision with 4 significant figures accuracy and Humlíček's code when run using single precision is also included in the Table 4.

**Table 3**: Details of the four different cases used for speed comparison of the present *Fortran* version with other competitive *Fortran* codes. For all worked cases, a grid of 2,840,071 input values for $z$ is used with 71 values for $y$ and 40001 values for $x$.

|  | Case 1 | Case 2 | Case 3 | Case 4 |
|---|---|---|---|---|
| $y$ uniformly spaced on the logarithmic scale | $[10^{-5}, 10^{5}]$ | $[10^{-20}, 10^{4}]$ | $[10^{-5}, 10^{5}]$ | $[10^{-20}, 10^{6}]$ |
| $x$ uniformly spaced on the linear scale | $[-500, 500]$ | $[-200, 200]$ | $[-10, 10]$ | Randomly generated to satisfy $|z|^2 \leq 36$ |

For each of the three codes (Weideman ($N=16$ and $N=32$), Poppe & Wijers, and Humlíček (double and single precision)) used for comparison with the present *Fortran* version in Table 4, the order of magnitude of the maximum relative error observed with the computation of each of the real and imaginary parts of the function is given. The reported errors are calculated using results from the present *Fortran* version with 13 significant digits as reference values. The reported errors are *underlined* whenever the observed value remarkably exceeds the claimed accuracy of the particular algorithm. Errors resulting from using the present version of the code for a number of significant figures less than 13 have been verified to be consistent with the claimed accuracy and therefore, they are not included in the Table for brevity.

The results in Table 4 indicate that the present *Fortran* version with 13 significant digits is faster than Algorithm 680 for all cases except for the second case. However, in none of the four worked cases did Algorithm 680 consistently and adequately satisfy its claimed accuracy for the real part of the function. This is due to the accuracy failure of the code for calculating the real part of the function around $x=6.3$ and small values of $y$ as pointed out in [Zaghloul and Ali 2011].

Compared to Weideman's algorithm ($N=16$), *Faddeyeva_v2(z,6)* is faster for the first 2 cases though slower for cases 3 and 4. A similar result is obtained when comparing

Weideman's Algorithm ($N$=32) with *Faddeyeva_v2(z,12)*. However, in none of the four worked cases did Weideman's algorithm consistently and adequately satisfy its claimed accuracy for the real part of the function. Apparently Weideman's algorithm suffers a loss of accuracy problem in the computations of the real part of the function for small values of *y* as well. Even for cases 1 and 3 where *y* changes between $10^{-5}$ and $10^{5}$, a maximum relative error in the order of $10^{-6}$ is obtained for the real part of the function while the claimed accuracy is 12 significant digits.

Comparison of the low accuracy (4 significant figures) version with results from Humlíček's algorithm indicates that Humlíček's algorithm is slightly faster for case 1 using double precision calculations and slightly slower for the same case using single precision calculations. For case 2, Humlíček's algorithm is faster by a factor <1.6 for double precision calculations and a factor <1.2 for single precision calculations. For case 3, Humlíček's algorithm is faster by a factor slightly >4 for double precision calculations and a factor <4 for single precision calculations while for case 4 these factors reduce in order to slightly >3 and slightly <2.7. The code also consistently fails to satisfy its claimed accuracy in the computation of the real part of the function for cases 2 and 4 as it suffers from a loss of accuracy problem near the real axis as pointed out in [Zaghloul and Ali 2011]. In addition, it has to be noted that the borders used herein for the application of the asymptotic expression in Eq (2) are very strict and extremely conservative in terms of satisfying the claimed accuracy over a very wide (all real axis and $y \in [10^{-40}, 10^{40}]$) and very fine grid. A bit of relaxation in these borders could marginally improve the performance. For example, reducing the border of applying the asymptotic expression in Eq (2) for the case of 4 significant figures from $|z|^2 \geq 107$ to $|z|^2 \geq 64$ will reduce the execution time by 25% for case 3 while maintaining the claimed accuracy for the tested case. It is also worth noting that both Algorithm 680 and Humlíček's algorithm are of a single fixed accuracy (less flexible) and lose their accuracy in some regions of the computational domain that may or may not be of practical interest depending on the type of application under consideration. Similarly, Weideman's algorithm also consistently fails to satisfy its claimed accuracy for both ($N$=16 and $N$=32).

A simple improvement to Humlíček's w4 algorithm [Zaghloul 2015] corrects the accuracy problem near the real axis. This has been implemented in the *Fortran* code for the case of 4 significant digits and produces computational efficiency similar to methods currently used for practical applications. Table 5 gives timings for this revised method and shows a clear improvement over the original *Faddeyeva_v2(z,4)* implementation with times very much in line with the basic Humlíček's routine but without the loss of accuracy.

**Table 4**: Average time per evaluation of the present *Fortran* version (requesting different numbers of significant figures) compared with other competitive algorithms. Calculations have been performed using *Intel Fortran Compiler version 11.1.038*.

| Algorithm | Average time per evaluation (μs) | | | | Claimed accuracy and comments |
|---|---|---|---|---|---|
| | Case 1 | Case 2 | Case 3 | Case 4 | |
| *Faddeyeva_v2(z, 13)* | 0.0623 | 0.1726 | 0.2328 | 0.3001 | ▪ *13 significant digits* |
| *Faddeyeva_v2(z, 12)* | 0.0445 | 0.1190 | 0.2166 | 0.2883 | ▪ *12 significant digits* |
| *Faddeyeva_v2(z, 11)* | 0.0336 | 0.0856 | 0.2027 | 0.2786 | ▪ *11 significant digits* |
| *Faddeyeva_v2(z, 10)* | 0.0274 | 0.0672 | 0.1985 | 0.2786 | ▪ *10 significant digits* |
| *Faddeyeva_v2(z, 9)* | 0.0226 | 0.0518 | 0.1827 | 0.2656 | ▪ *9 significant digits* |
| *Faddeyeva_v2(z, 8)* | 0.0202 | 0.0438 | 0.1814 | 0.2656 | ▪ *8 significant digits* |
| *Faddeyeva_v2(z, 7)* | 0.0170 | 0.0326 | 0.1658 | 0.2523 | ▪ *7 significant digits* |
| *Faddeyeva_v2(z, 6)* | 0.0151 | 0.0255 | 0.1609 | 0.2523 | ▪ *6 significant digits* |
| *Faddeyeva_v2(z, 5)* | 0.0141 | 0.0217 | 0.1488 | 0.2382 | ▪ *5 significant digits* |
| *Faddeyeva_v2(z, 4)* | 0.0136 | 0.0207 | 0.1392 | 0.2252 | ▪ *4 significant digits* |
| Single Precision *Faddeyeva_v2(z, 4)* | 0.0112 | 0.0156 | 0.0953 | 0.1544 | ▪ *4 significant digits* |
| Poppe & Wijers [1990] | 0.0863 | 0.1245 | 0.2871 | 0.5711 | ▪ *13 significant digits* |
| $\delta_{V,max}$ | $\times 10^{-9}$ | $\times 10^{+1}$ | $\times 10^{-9}$ | $\times 10^{-1}$ | ▪ Loss of claimed accuracy |
| $\delta_{L,max}$ | $\times 10^{-12}$ | $\times 10^{-12}$ | $\times 10^{-13}$ | $\times 10^{-13}$ | noticed for $y \leq 10^{-5}$ |
| Humlíček [1982] | 0.0118 | 0.0132 | 0.0317 | 0.0722 | ▪ *4 significant digits* |
| $\delta_{V,max}$ | $\times 10^{-4}$ | $\times 10^{+1}$ | $\times 10^{-4}$ | $\times 10^{+1}$ | ▪ Loss of claimed accuracy |
| $\delta_{L,max}$ | $\times 10^{-4}$ | $\times 10^{-4}$ | $\times 10^{-4}$ | $\times 10^{-4}$ | noticed for $y \leq 10^{-7}$ |
| Single precision | 0.0122 | 0.0135 | 0.0255 | 0.0574 | ▪ *4 significant digits* |
| $\delta_{V,max}$ | $\times 10^{-4}$ | $\times 10^{+1}$ | $\times 10^{-4}$ | $\times 10^{+1}$ | ▪ Same loss of claimed |
| $\delta_{L,max}$ | $\times 10^{-4}$ | $\times 10^{-4}$ | $\times 10^{-4}$ | $\times 10^{-4}$ | accuracy noticed above |
| Weideman [1994] *N*=16 terms | 0.0688 | 0.0687 | 0.0687 | 0.0688 | ▪ <u>*6 significant digits*</u> |
| $\delta_{V,max}$ | $\times 10^{+3}$ | $\times 10^{+17}$ | $\times 10^{0}$ | $\times 10^{+9}$ | ▪ Failure to fully satisfy the |
| $\delta_{L,max}$ | $\times 10^{-5}$ | $\times 10^{-5}$ | $\times 10^{-5}$ | $\times 10^{-6}$ | claimed accuracy for the real part for all 4 cases |
| *N*=32 terms | 0.1321 | 0.1323 | 0.1327 | 0.1323 | ▪ <u>*12 significant digits*</u> |
| $\delta_{V,max}$ | $\times 10^{-6}$ | $\times 10^{+9}$ | $\times 10^{-6}$ | $\times 10^{+3}$ | ▪ Failure to fully satisfy the |
| $\delta_{L,max}$ | $\times 10^{-12}$ | $\times 10^{-12}$ | $\times 10^{-12}$ | $\times 10^{-12}$ | claimed accuracy for the real part for all 4 cases |

**Table 5**: Average time per evaluation and accuracy of the present *Fortran* version *Faddeyeva_v2(z, 4)* implementing a proposed reform for Humlíček's approximation for the case of 4 significant figures [Zaghloul 2015]. Calculations have been performed using *Intel Fortran Compiler version 11.1.038*.

| Algorithm | Average time per evaluation (μs) | | | | Claimed accuracy and comments |
|---|---|---|---|---|---|
| | Case 1 | Case 2 | Case 3 | Case 4 | |
| Double Precision *Faddeyeva_v2(z, 4)* $\delta_{V,max}$ $\delta_{L,max}$ | 0.0119 $\times 10^{-4}$ $\times 10^{-4}$ | 0.0147 $\times 10^{-4}$ $\times 10^{-4}$ | 0.0292 $\times 10^{-4}$ $\times 10^{-4}$ | 0.0680 $\times 10^{-4}$ $\times 10^{-4}$ | ▪ *4 significant digits* |
| Single Precision *Faddeyeva_v2(z, 4)* $\delta_{V,max}$ $\delta_{L,max}$ | 0.0101 $\times 10^{-4}$ $\times 10^{-4}$ | 0.0128 $\times 10^{-4}$ $\times 10^{-4}$ | 0.0264 $\times 10^{-4}$ $\times 10^{-4}$ | 0.0636 $\times 10^{-4}$ $\times 10^{-4}$ | ▪ *4 significant digits* |

## 6. SUMMARY AND CONCLUSIONS

Efficiency improvements to Algorithm 916 are reported and discussed. The efficiency of the algorithm, when run at its highest accuracy is improved by a factor close to three. A more effective accuracy vs efficiency trade-off scheme is also implemented which improves the efficiency by an additional factor of three. The current version of the code is provided in *Matlab* and *Scilab*. In addition a *Fortran* translation has been implemented to meet the needs of real-world problems where the very large number of function evaluations would require the use of a compiled language.

## ACKNOWLEDGMENTS

The author would like to acknowledge valuable and insightful comments and suggestions received from the reviewers, associate editor and the algorithm editor.